\documentclass{emulateapj}
\usepackage{apjfonts}

\newcommand{\Msun}      {\mbox{$\rm\,M_{\mathord\odot}$}}

\begin{document}

\title{The Low Quiescent X-Ray Luminosity of the Transient X-Ray 
Burster EXO 1747--214}

\author{John A. Tomsick\altaffilmark{1},
Dawn M. Gelino\altaffilmark{2},
Philip Kaaret\altaffilmark{3}}

\altaffiltext{1}{Center for Astrophysics and Space Sciences, Code
0424, University of California at San Diego, La Jolla, CA,
92093 (e-mail: jtomsick@ucsd.edu)}

\altaffiltext{2}{Michelson Science Center, California Institute of 
Technology, 770 South Wilson Avenue, MS 100-22, Pasadena, CA 91125}

\altaffiltext{3}{Department of Physics and Astronomy, University of
Iowa, Iowa City, IA 52242}

\begin{abstract}

We report on X-ray and optical observations of the X-ray burster
EXO~1747--214.  This source is an X-ray transient, and its only
known outburst was observed in 1984--1985 by the {\em EXOSAT}
satellite.  We re-analyzed the {\em EXOSAT} data to derive the
source position, column density, and a distance upper limit using
its peak X-ray burst flux.  We observed the EXO~1747--214 field
in 2003 July with the {\em Chandra X-ray Observatory} to search
for the quiescent counterpart.  We found one possible candidate 
just outside the {\em EXOSAT} error circle, but we cannot
rule out the possibility that the source is unrelated to 
EXO~1747--214.  Our conclusion is that the upper limit on the 
unabsorbed 0.3--8 keV luminosity is $L < 7\times 10^{31}$ 
erg~s$^{-1}$, making EXO~1747--214 one of the faintest neutron 
star transients in quiescence.  We compare this luminosity upper 
limit to the quiescent luminosities of 19 neutron star and 14 
black hole systems and discuss the results in the context of the 
differences between neutron stars and black holes.  Based on the 
theory of deep crustal heating by Brown and coworkers, the 
luminosity implies an outburst recurrence time of $>$1300 yr 
unless some form of enhanced cooling occurs within the neutron 
star.  The position of the possible X-ray counterpart is consistent 
with three blended optical/IR sources with $R$-magnitudes between 
19.4 and 19.8 and $J$-magnitudes between 17.2 and 17.6.  One of 
these sources could be the quiescent optical/IR counterpart of 
EXO~1747--214.

\end{abstract}

\keywords{accretion, accretion disks --- stars: neutron ---
stars: individual (EXO 1747--214) --- X-rays: stars --- X-rays: general}

\section{Introduction}

Observations of transient X-ray binary systems in quiescence
have greatly advanced studies of compact objects over the past
decade.  While both black hole and neutron star systems can 
approach or possibly even exceed their Eddington luminosities 
during outbursts, their quiescent X-ray luminosities can be 
factors of $>$$10^{7}$ lower than their peak luminosities.  
Quiescent optical observations allow for the measurement of 
compact object masses as the optical companion's radial velocity 
curves can be measured as well as the binary inclination via 
studies of ellipsoidal modulations
\citep[][and references therein]{cc03}.  Such measurements 
have been especially important for confirming that there is a 
population of compact objects with masses that are too high
($>$3\Msun) to be neutron stars so that these objects are
very likely black holes.  In the future, mass measurements of
neutron star transients may lead to constraints on the equation
of state for matter at high densities \citep{lp04}.  

Quiescent X-ray observations are also interesting for neutron
star and black hole systems.  For many neutron star systems, 
a thermal component is seen in the quiescent X-ray spectrum
that is likely blackbody emission from the neutron star 
surface.  Measurements of the evolution of this component 
allow for constraints on the thermal properties of the 
neutron star crust and core \citep{wijnands04_cool}.  The 
core temperature is thought to be set by the accretion history 
over a time period of 10,000 years \citep{bbr98,colpi01}.  
Quiescent X-ray spectra of black hole systems do not appear 
to show a blackbody component, which is one reason they tend 
to be fainter than neutron star systems.  It has also been 
argued that, in quiescence, black holes are radiatively less 
efficient than neutron star systems, which has been taken as 
evidence that a large fraction of the accreted matter is 
advected across the black hole event horizon 
\citep{ngm97,mcclintock03}.

With sensitive X-ray missions like the {\em Chandra X-ray 
Observatory} and the {\em X-ray Multi-Mirror Mission 
(XMM-Newton)}, it has been possible to observe many more 
neutron star and black hole systems in quiescence.  As more 
systems are observed, it has been found that many of the 
neutron star systems are fainter than the $10^{32-34}$ 
erg~s$^{-1}$ range usually considered as the typical range 
for neutron stars in quiescence
\citep{campana02,wijnands05,tomsick04_2123,jwv04}.
Thus, it is important to obtain the largest samples possible
to avoid any selection biases when comparing the quiescent
properties of black holes and neutron stars.

In this paper, we use archival data as well as our recent
observations to study the neutron star X-ray transient
EXO 1747--214.  This system was discovered during the
{\em EXOSAT} Galactic plane scans in 1984 June
\citep{warwick88,parmar85} and was detected again
by {\em EXOSAT} in 1985 April at about the same flux
level \citep{magnier89,parmar85}.  These are the only two 
known detections of the source, so that it has apparently 
been in quiescence for 20 years.  An X-ray burst 
was detected in 1985, demonstrating that the system 
definitely harbors a neutron star \citep{magnier89}.  
Although {\em EXOSAT} provided a relatively good 
$15^{\prime\prime}$ position for the source, and the
extinction along the line of sight to the source
($l = 7^{\circ}.00, b = +2^{\circ}.95$) is fairly
low, the source has not been optically identified.
The primary goal of this work is to use a {\em Chandra}
observation of moderate length to detect the source
in quiescence and to search for an optical counterpart.
We report on our search below, including a constraint
on the quiescent luminosity and a comparison between
the Eddington-scaled luminosity of EXO~1747--214 and
those of 19 neutron star and 14 black hole systems.

\section{Analysis and Results}

The data we used for this work come from the X-ray, optical, 
and infrared observations listed in Table~\ref{tab:obs}.  First, 
we went back to the discovery observations made by {\em EXOSAT} 
in 1984--1985.  During this time, the source was bright in X-rays, 
and we used the {\em EXOSAT} images to re-derive the source 
position.  We also re-analyzed the data from the {\em EXOSAT} 
proportional counters to obtain information about the energy 
spectrum and the detected X-ray burst.  The main focus of this 
paper is to search for the quiescent X-ray counterpart.  For 
this purpose, we used the data from our 2003 {\em Chandra} 
observation.  Finally, we obtained optical and infrared images 
to allow for optical or IR identifications of any detected X-ray 
sources.  Here, we describe our analysis of the data from these 
observations and our results.

\subsection{Source Position from {\em EXOSAT}}

\begin{figure}
\centerline{\includegraphics[width=0.50\textwidth]{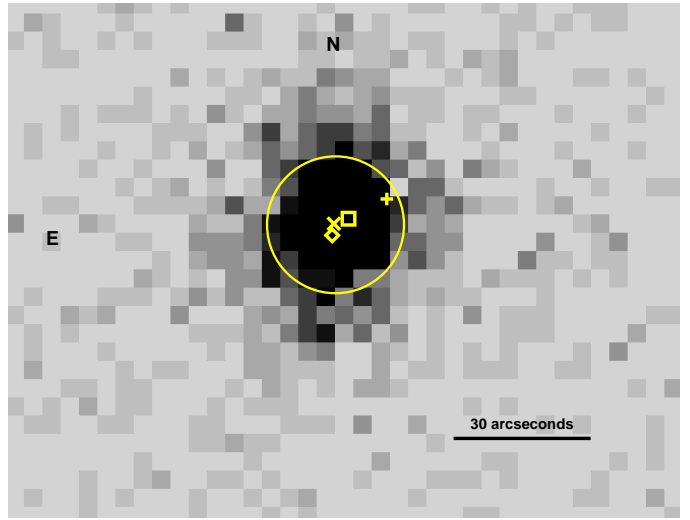}}
\caption{The 0.05--2.5 keV image of EXO~1747--214 in outburst.
The data come from the {\em EXOSAT} CMA instrument.  The 
observation occurred on 1985 April 11, and the exposure time 
was 9,171~s.  The $15^{\prime\prime}$ radius circle shows 
the 90\% confidence (including systematic pointing errors) 
{\em EXOSAT} error region for EXO~1747--214.  The {\em cross}, 
the {\em diamond}, and the {\em square} show the positions 
derived from data taken with the three CMA filters on 1985 
April 11.  The {\em plus} marks the position obtained from
the observation made on 1985 April 8.\label{fig:image_exosat}}
\end{figure}

{\em EXOSAT} \citep{taylor81} observations of EXO~1747--214 occurred 
in 1985 on April 8 and 11.  The most precise position for this source 
comes from data obtained with the low energy imaging telescope (LE1) 
and the Channel Multiplier Array (CMA1) instrument, which has a 
bandpass of 0.05--2.5 keV.  However, as the {\em EXOSAT}/CMA 
catalog\footnote{The {\em EXOSAT}/CMA catalog is part of the ``Master X-Ray 
Catalog'' at http://heasarc.gsfc.nasa.gov/W3Browse/all/xray.html} does not 
give a single unique position for EXO~1747--214, we re-analyzed the CMA
data to determine the position.  During the short April 8 observation, 
the Thin Lexan filter was inserted, and EXO~1747--214 was $\sim$6$^{\prime}$
off-axis.  On April 11, the source was on-axis, and three filters 
were used at different times:  Thin Lexan, with the highest effective 
area ($A_{\rm eff}$) over the full bandpass; Al/Parylene, with reduced 
$A_{\rm eff}$ below 0.3 keV; and Boron, with two bands of $A_{\rm eff}$ 
peaking at 0.15 keV and 1 keV \citep{dekorte81}.  With the highest
$A_{\rm eff}$ and the longest exposure time (see Table~\ref{tab:obs}), 
the image obtained with the Thin Lexan filter provides the best
statistics, and this image is shown in Figure~\ref{fig:image_exosat}.
Nearly all of the source counts are contained in a 24-by-24 pixel
region centered on the source.  For the image shown in 
Figure~\ref{fig:image_exosat}, there are 1331 counts in this region, 
and using number of counts per pixel from the surrounding, source-free, 
region, we estimate that there are 104 background counts in a
24-by-24 pixel region.  For each of the four images, we calculated 
the centroid using the counts from a 24-by-24 pixel region centered 
on the source, and these centroids are shown in Figure~\ref{fig:image_exosat}.
The position on April 8 is $\sim$10$^{\prime\prime}$ away from the
other three positions, and this is probably due to the fact that 
the source was off-axis on April 8 so that the position is not as
reliable.  We combined the data for the three on-axis images and
recalculated the centroid to determine a best source position of
R.A. = $17^{\rm h}50^{\rm m}25^{\rm s}\!.8$, 
decl. = $-21^{\circ}25^{\prime}22^{\prime\prime}$ (equinox J2000).
The {\em EXOSAT}/CMA catalog indicates that the 90\% confidence 
error radius is $15^{\prime\prime}$, and the corresponding error 
circle is shown in Figure~\ref{fig:image_exosat}.  The best position 
that we derive is $4^{\prime\prime}.5$ from the median of the 
positions given in the {\em EXOSAT}/CMA catalog, and it is 
$2^{\prime\prime}.8$ from the position quoted in \cite{gottwald91}.  
Thus, all these positions are well within the $15^{\prime\prime}$ 
error circle.

\subsection{Column Density and Distance Upper Limit from {\em EXOSAT}}

On 1985 April 11, EXO 1747--214 was also observed with the 
{\em EXOSAT} medium energy (ME) instrument, which consists
of eight proportional counters.  The observation lasted for
more than 6 hours, and, as reported in \cite{magnier89}, 
included one type I X-ray burst.  The original analysis by
\cite{magnier89} reported that the 1--20 keV spectrum of the
persistent emission was well-described by a power-law with
interstellar absorption with a column density of 
$N_{\rm H} = 1\times 10^{22}$ cm$^{-2}$.  To check this, 
we obtained the ME spectrum from the {\em EXOSAT} archive.
The details of the spectral extraction and background 
subtraction are provided 
on-line\footnote{http://heasarc.gsfc.nasa.gov/W3Browse/exosat/me.html}.
When the EXO 1747--214 spectrum was extracted, the data from 
the time of the X-ray burst was removed, leaving the spectrum
of the persistent emission.

\begin{figure}
\centerline{\includegraphics[width=0.50\textwidth]{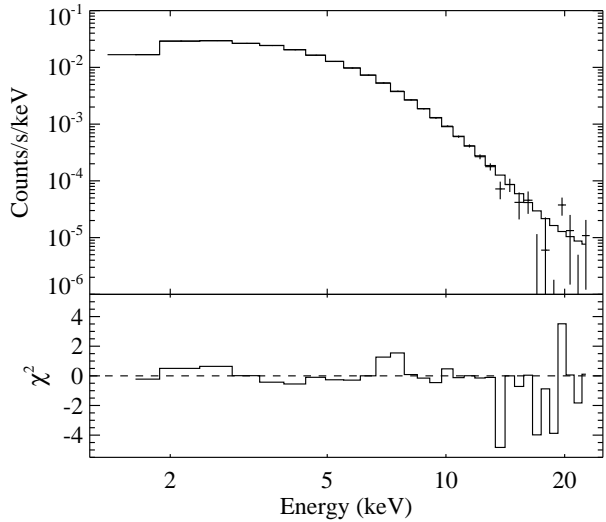}}
\vspace{-1.2cm}
\caption{The {\em EXOSAT} ME energy spectrum from a 22,450~s
observation made on 1985 April 11.  The spectrum is fitted
with a Comptonization (``comptt'') model and interstellar
absorption with $N_{\rm H}$ fixed at the Galactic value of
$4\times 10^{21}$ cm$^{-2}$.  The fitted ``comptt'' parameter
values include: The temperature of the ``seed'' photons, 
$0.34\pm 0.02$ keV; the plasma temperature, $2.17\pm 0.06$ keV;
and the optical depth of the plasma, $7.0\pm 0.2$.  The data and
model are shown in the top panel, and the residuals are shown in 
the bottom panel.  We measure a 1--20 keV flux of 
$1.45\times 10^{-9}$ erg~cm$^{-2}$~s$^{-1}$.\label{fig:spectrum}}
\end{figure}

We used the spectral fitting package XSPEC version 11.3.1t to 
perform least-squares fits to the ME spectrum, and our results 
do not confirm those of \cite{magnier89}.  A power-law with 
(interstellar) absorption gives a very poor fit to the spectrum 
with $\chi^{2}/\nu = 498/28$.  The reason for the poor fit is
that the spectrum drops more rapidly at high energies than a
power-law model, indicating the presence of a cutoff.  Models
that include a cutoff greatly improve the fit.  An exponentially
cutoff power-law (``cutoffpl'' in XSPEC) with absorption gives 
$\chi^{2}/\nu = 68/27$ and a much lower column density than
$1\times 10^{22}$ cm$^{-2}$.  A Comptonization model (``comptt'' 
in XSPEC) with absorption gives a statistically acceptable fit 
with $\chi^{2}/\nu = 26/26$.  The spectrum fitted with a 
Comptonization model is shown in Figure~\ref{fig:spectrum}.  
With this fit, the column density, using \cite{ag89} abundances 
and \cite{bm92} cross-sections, is $N_{\rm H} = 
(1.9^{+2.8}_{-1.9})\times 10^{21}$ cm$^{-2}$ (90\% confidence 
errors), which is consistent with the Galactic value of 
$4.0\times 10^{21}$ cm$^{-2}$ \citep{dl90}.  We fixed $N_{\rm H}$ 
to the Galactic value and re-fitted the 1--20 keV spectrum.  The 
Comptonization parameters from this fit are given in the caption 
of Figure~\ref{fig:spectrum}.  Below, we assume that the 
interstellar column density is at the Galactic value.

We use the fact that the peak X-ray burst flux must be below the
Eddington luminosity to derive an upper limit on the distance 
to EXO 1747--214.  To derive the peak flux, we use the X-ray 
burst spectral results from \cite{magnier89} and the peak 
1--8 keV count rate of 2343 c/s, which we obtain from our 
re-analysis of the {\em EXOSAT} data.  Fitting spectra around the 
peak of the burst with 0.5~s time resolution with a blackbody 
model, \cite{magnier89} find that the peak burst temperature is 
$\sim$2 keV.  After fixing the temperature to 2~keV and assuming 
$N_{\rm H} = 4\times 10^{21}$ cm$^{-2}$, we adjusted the blackbody 
normalization to give the peak 1--8 keV count rate.  This leads 
to an unabsorbed 1--20 keV flux of $2.7\times 10^{-8}$ 
erg~cm$^{-2}$~s$^{-1}$.  For radius expansion bursts from neutron 
stars in globular clusters (i.e., Eddington limited bursts from 
neutron stars at known distances), \cite{kuulkers03} found an 
average peak luminosity of $3.8\times 10^{38}$ erg~s$^{-1}$.  
From this luminosity and the peak flux of the EXO 1747--214 burst, 
we derive a distance upper limit of $d < 11$ kpc.  

\subsection{A Search for the Quiescent X-Ray Counterpart with {\em Chandra}}

We obtained a 24.5 ks {\em Chandra} observation of the EXO 1747--214
field on 2003 July 31 (Observation ID 3802).  We used the Advanced 
CCD Imaging Spectrometer \citep[ACIS,][]{garmire03} with the target 
position placed on one of the back-illuminated ACIS chips (ACIS-S3).  
For our observation, the instrument was in the ``VFAINT'' mode, 
which provides the maximum information about each event and thus, 
the best rejection of background events.  We started with the 
``level 1'' event list produced by the standard data processing 
with ASCDS version 6.13.4 and performed further processing using 
the {\em Chandra} Interactive Analysis of Observations (CIAO) 
version 3.2 software and Calibration Data Base (CALDB) version 
3.0.1.  We used the CIAO routine {\em acis\_process\_events} to 
obtain a ``level 2'' event list and image, and we searched for 
sources using this image.

\begin{figure}
\centerline{\includegraphics[width=0.50\textwidth]{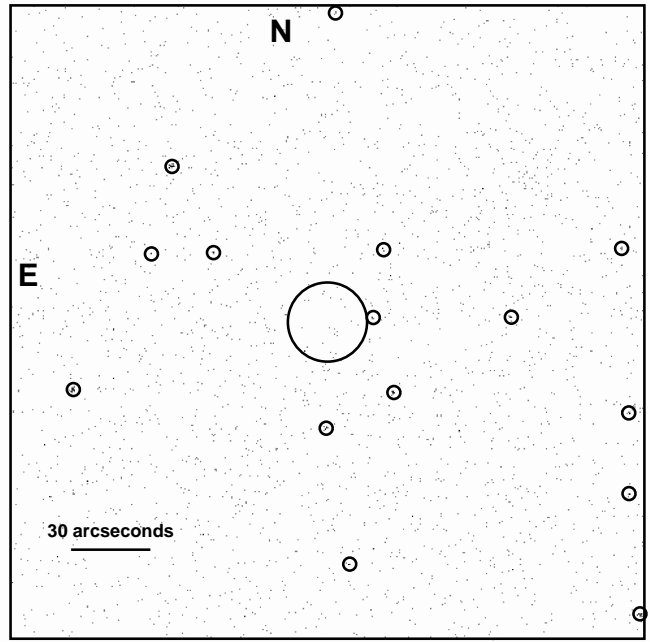}}
\caption{The 0.3--8 keV {\em Chandra}/ACIS image of the 
EXO 1747--214 field.  The exposure time is 24.5 ks, and the
image size is $4^{\prime}$-by-$4^{\prime}$.  The 
$15^{\prime\prime}$ radius circle shows the 90\% confidence 
{\em EXOSAT} error circle for EXO~1747--214.  The other 15 
circles mark the detected {\em Chandra} sources.
\label{fig:image_chandra}}
\end{figure}

For our source search, we focused on a $4^{\prime}$-by-$4^{\prime}$
square region of the ACIS-S3 chip centered on the {\em EXOSAT} 
position.  We restricted the energy range to 0.3--8 keV and used 
the CIAO routine {\em wavdetect} \citep{freeman02} to search for
sources.  Based on the image size of 488-by-488 pixels and our 
detection threshold of $4.2\times 10^{-6}$, we would expect to 
detect $\sim$1 spurious source.  We detected 15 sources ranging 
from 3 to 24 counts.  Using the background estimates generated by 
{\em wavdetect} and Poisson statistics, the 3 and 4 count sources 
are detected at significances of $\sim$3.5-$\sigma$ and 
$\sim$4.1-$\sigma$, respectively.  Figure~\ref{fig:image_chandra} 
shows the {\em Chandra} image with the 15 sources circled.  The 
$15^{\prime\prime}$ 90\% confidence {\em EXOSAT} error circle is 
also shown.  None of the {\em Chandra} sources fall within the 
error circle, and only one of the sources lies close enough 
($17^{\prime\prime}.5$ away) to be considered a possible quiescent 
counterpart to EXO 1747--214.  The position of this possible 
counterpart, which is one of the sources with 4 counts, is 
R.A. = $17^{\rm h}50^{\rm m}24^{\rm s}\!.52$, decl. = 
$-21^{\circ}25^{\prime}19^{\prime\prime}\!.9$ (equinox J2000, 
$0^{\prime\prime}\!.6$ uncertainty).  Given the density of 
sources that we detect (15 sources in a 16 arcmin$^{2}$ region), 
the probability of finding a source within $17^{\prime\prime}.5$ 
of the best EXO 1747--214 position by chance is 19\%.  Thus, the 
possible counterpart could be an unrelated X-ray source.

Although the 3-$\sigma$ detection threshold for the {\em EXOSAT}
error region is close to 3 counts, considering the possible
4 count source just outside the {\em EXOSAT} error circle, 
the most concise statement of our results is that the number of 
0.3--8 keV counts from the quiescent EXO 1747--214 counterpart 
is $\leqslant$4.  This corresponds to a limit on the 0.3--8 keV 
count rate of $\leqslant$$1.63\times 10^{-4}$ c/s.  To determine 
what this implies in terms of the source flux and luminosity, 
we used the CIAO tools to produce an ACIS response matrix 
for this observation.  To calculate the flux upper limit, 
we assumed $N_{\rm H} = 4\times 10^{21}$ cm$^{-2}$ and 
several different spectral models that are similar to
the typical X-ray spectra of quiescent neutron star
systems, including a thermal model, a non-thermal model, 
and a two-component model.  For the thermal model, we
used a neutron star atmosphere model
\citep[``nsa'' in XSPEC,][]{zps96} and assumed a source
distance of 11~kpc (the upper limit), a neutron star mass 
of 1.4\Msun, and a neutron star radius of 10 km.  The only
other free parameter in the model is the temperature
($kT_{\rm eff}$).  We adjusted $kT_{\rm eff}$ to make the
predicted count rate match the upper limit given above, 
and this occurs at 63~eV.  For this model, the absorbed
0.3--8 keV flux is $5.6\times 10^{-16}$ erg~cm$^{-2}$~s$^{-1}$ 
and the unabsorbed flux is $4.9\times 10^{-15}$ 
erg~cm$^{-2}$~s$^{-1}$.  Non-thermal power-law models with 
photon indeces of $\Gamma = 2$ and 3 give upper limits on 
the unabsorbed flux of $2.5\times 10^{-15}$ and $3.6\times 
10^{-15}$ erg~cm$^{-2}$~s$^{-1}$, respectively.  We also 
considered two-component, nsa plus power-law, models with 
$kT_{\rm eff} = 55$~eV along with the same nsa parameters
as above.  For photon indeces of $\Gamma = 1$ and 2, the
upper limits on the unabsorbed flux are $4.2\times 10^{-15}$ 
and $3.9\times 10^{-15}$ erg~cm$^{-2}$~s$^{-1}$, respectively. 
In summary, the highest flux is obtained using the pure 
thermal model.  Using this flux ($4.9\times 10^{-15}$ 
erg~cm$^{-2}$~s$^{-1}$) along with the fact that $d < 11$~kpc, 
we conclude that $L < 7\times 10^{31}$ erg~s$^{-1}$ (0.3--8 keV).

\subsection{Optical and Infrared Observations}

\begin{figure}
\centerline{\includegraphics[width=0.45\textwidth]{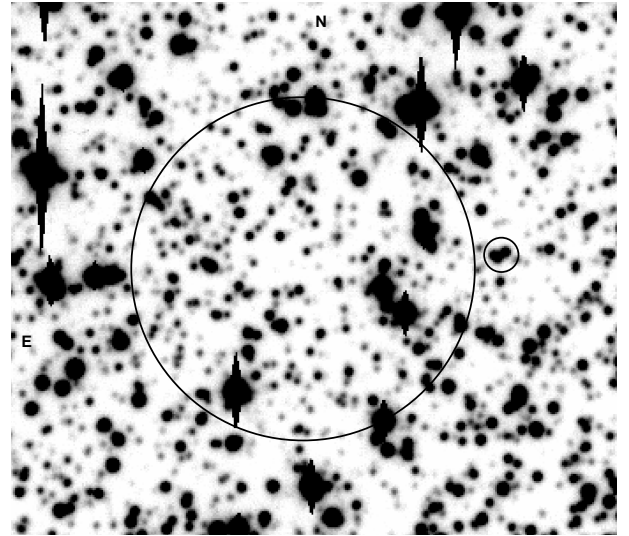}}
\caption{A 600~s $R$-band image taken using the ESI instrument
at Keck Observatory.  The large circle sets the scale and is
the $15^{\prime\prime}$ 90\% confidence {\em EXOSAT} error 
circle for EXO~1747--214.  The smaller circle marks the 
location of a four count {\em Chandra} source.  Its position
is consistent with any of the three $R = 19.4$--19.8 optical 
sources in the smaller circle.\label{fig:image_keck}}
\end{figure}

A couple of months after the {\em Chandra} observation, 
we obtained a deep $R$-band image of the EXO 1747--214 
field at Keck Observatory.  For the single 600~s image, 
we performed bias subtraction and flat-fielding in the 
standard manner using IRAF\footnote{IRAF (Image Reduction 
and Analysis Facility) is distributed by the National 
Optical Astronomy Observatories, which are operated by 
the Association of Universities for Research in Astronomy, 
Inc., under cooperative agreement with the National 
Science Foundation.}.  To register the image, we identified 
six of the stars in the image with stars in the USNO-B
catalog, and used IRAF routines to calculate the sky
projection.  For magnitude calibration, we obtained
observations of \cite{landolt92} standards PG 1657+078
and PG 0231+051 before and after observing the 
EXO 1747--214 field and performed aperture photometry
for the standard stars as well as for several stars in
the EXO 1747--214 field.  Figure~\ref{fig:image_keck}
shows the central part of the Keck image, including
the EXO 1747--214 error circle and the possible 
{\em Chandra} counterpart described above.  The 
possible counterpart is positionally coincident
with three blended optical sources as shown in the
figure.  The three optical sources are approximately
the same brightness, and all three are in the 
$R = 19.4$--19.8 range.  

We also obtained an infrared $J$-band image of the 
EXO 1747--214 field in 2002.  We used the 4 meter 
telescope at Cerro Tololo Inter-American Observatory 
(CTIO) along with the Optical System for Imaging and 
Low-Resolution Integrated Spectroscopy (OSIRIS)
instrument.  We detect the three optical sources
discussed above, and all three are in the 
$J = 17.2$--17.6 range.

While this positional coincidence certainly does not 
prove that this source is the EXO 1747--214 counterpart, 
it is notable that the properties of the optical sources 
are consistent with what might be expected for the 
EXO 1747--214 counterpart.  The sources appear to
be point-like as opposed to being extended as might be
the case if they are galaxies.  Also, the magnitude
range is not unreasonable for a quiescent neutron star
LMXB at a few kpc.  For $R = 19.6$, $A_{R} = 1.7$, and 
$d = 3$~kpc, the absolute magnitude is $M_{R} = 5.5$, 
which is consistent with a K3V or K4V star.  In 
addition, a $J$-magnitude of 17.4 and $A_{J} = 0.6$ 
indicate a de-reddened $R-J$ color of 1.1, which is 
also consistent with main sequence K-type star.  If the 
possible counterparts have spectra consistent with a
those of main sequence stars, then they are not at a 
distance that is much more than a few kpc.  For example, 
the $R$-band measurement and a distance of 11~kpc imply 
$M_{R} = 2.7$ and a late-F spectral type, but the $R-J$ 
colors are not consistent with an F-type star (even 
considering the range of measured $R$ and $J$-magnitudes).

\section{Discussion}

\subsection{Neutron Star/Black Hole Comparison}

The upper limit on the EXO 1747--214 X-ray luminosity
of $7\times 10^{31}$ erg~s$^{-1}$ (0.3--8 keV) places 
it among the faintest of the quiescent neutron star
systems.  One reason that faint neutron star systems
are of interest is that claims for the existence of
black hole event horizons rely on the faintness of 
black hole systems relative to neutron star systems
\citep{ngm97,menou99,garcia01,mcclintock03}.  The 
faintness of the black hole systems may be due to 
advection of accretion energy across the event horizon 
as occurs in Advection-Dominated Accretion Flow (ADAF)
models \citep[][and references therein]{ngm97} or to 
the lack of the thermal component that is often present 
for quiescent neutron star systems.  In either case, 
one interpretation for the difference is that neutron 
stars have a solid surface while black holes do not.

To place our EXO~1747--214 luminosity upper limit in 
the context of this discussion, we compiled a list of
14 black hole and 20 (including EXO~1747--214) neutron 
star transients with constraints on the quiescent 
luminosity ($L_{\rm min}$).  The black hole list is 
complete in the sense that it includes all confirmed 
black hole systems for which sensitive quiescent X-ray 
observations have been made.  In all, there are 17 
dynamically confirmed black hole systems 
\citep{mr03,casares04,orosz04}.  Only GRS~1915+105, 
XTE~J1650--500, and GS~1354--64 do not have quiescent 
X-ray coverage.  We compiled the list of neutron star 
transients from the recent literature, including
papers where neutron star sub-populations were 
studied \citep[e.g.,][]{jonker04,tomsick04_2123,garcia01}.
Our list is nearly complete for field systems, but
we only included the best studied globular cluster
systems \citep[see][for additional globular cluster
sources]{heinke03}.

\begin{figure}
\centerline{\includegraphics[width=0.50\textwidth]{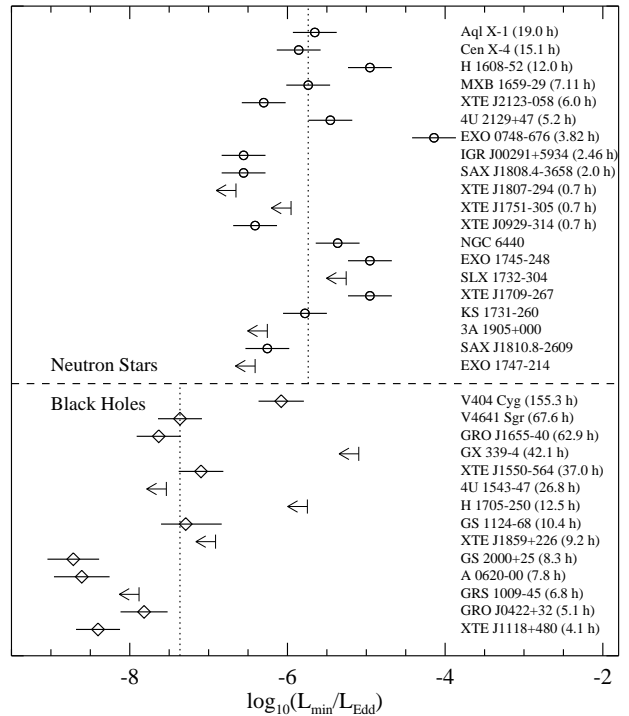}}
\caption{The Eddington-scaled minimum X-ray luminosities for
neutron star and black hole transients.  The bandpasses for the 
luminosities (which are unabsorbed) are nearly the same in each 
case (0.3--8 keV, 0.3--7 keV, or 0.5--10 keV).  The selection 
criteria and the description of how we calculated the values are 
described in detail in \S~3.1.  The black hole luminosities come 
from \cite{tomsick03_qbh} and references therein, and the 
neutron star luminosities come from the following: \cite{garcia01,jonker04,jwv04,campana02,campana05,tomsick04_2123,jc05,wijnands02,wijnands01,wijnands04,wijnands05,wijnands05_ter5,cackett05}.
For sources with measured orbital periods, $P_{\rm orb}$ is
given in parentheses after the source name.  The vertical 
dotted lines mark the median Eddington-scaled luminosities.
\label{fig:luminosity}}
\end{figure}

In previous work \citep[e.g.,][]{garcia01}, it has 
been argued that, for a black hole or neutron star 
system with a given orbital period ($P_{\rm orb}$), the 
Eddington-scaled mass accretion rate would be expected 
to be the same for black hole and neutron star systems.
If this is true, the Eddington-scaled X-ray luminosity 
is the best parameter for comparing the sources.  We 
calculated $L_{\rm min}/L_{\rm Edd}$ for our list of 
systems, and the values are shown in 
Figure~\ref{fig:luminosity}.  For the black hole systems, 
the luminosities come from \cite{tomsick03_qbh} and references
therein, and the black hole masses and errors on the 
masses come from \cite{cc03}.  The neutron star 
luminosities come from the references shown in the 
Figure~\ref{fig:luminosity} caption.
To determine the errors on $L_{\rm min}/L_{\rm Edd}$, 
we assume for all systems that the unabsorbed soft 
X-ray fluxes are measured to an accuracy of 40\% and that 
the distances are known to 25\%.  The assumed errors are 
not exactly correct for all sources, but they represent 
values that are typically quoted in the literature.  While 
the black hole errors also include a contribution from the
uncertainty in the black hole mass, we assume that all
the neutron stars have masses of 1.4\Msun.  The bandpasses
for the luminosities are nearly the same being 0.3--8~keV, 
0.3--7~keV, or 0.5--10~keV in all cases.  It is important 
to keep in mind that we are only quoting the X-ray 
luminosities here and that bolometric corrections could be 
important.  The sources in Figure~\ref{fig:luminosity} are 
ordered according to orbital period in cases for which 
$P_{\rm orb}$ is known.

The $\log_{10}({L_{\rm min}/L_{\rm Edd}})$ values shown in 
Figure~\ref{fig:luminosity} indicate that the black holes
are, on average, fainter.  For black holes, the median
luminosity is a factor of 42 lower than for neutron stars.
While part of this effect is due to the higher black hole
masses, the average black hole mass is 7.7\Msun, which is
only a factor of 5.5 higher than the neutron star masses
assumed.  Currently, the only clear overlap between the 
distributions is caused by the high luminosity of V404~Cyg, 
and as this source has the longest orbital period and has
an evolved optical companion, this may be due to a higher
quiescent mass accretion rate for V404~Cyg.  However, 
several of the neutron star systems, including EXO~1747--214,
have upper limits that are approaching the values seen for 
some of the other black hole systems.  It is important to 
determine how faint sources like EXO~1747--214 and
3A~1905+000 \citep{jc05} are (as well as their orbital 
periods) as the neutron star and black hole distributions 
could significantly overlap in contradiction to what is 
expected for accretion models that include black hole 
advection such as the ADAF model.

\subsection{Neutron Star Cooling}

One reason that EXO~1747--214 may be so faint is that, as
far as we know, it has been in quiescence for 20 years.
Based on the theory of deep crustal heating for transient
neutron stars \citep{bbr98}, the evolution of the X-ray 
flux in the thermal component between outbursts depends 
on the source's mass accretion (and thus outburst) history. 
\cite{rutledge02} show that for a long (13 year) outburst, 
the temperature of the crust can rise significantly, and
it can take years or decades (depending on the conductivity 
of the crust and the type of neutrino cooling that occurs) 
for the crust to reach thermal equilibrium.  At that point, 
the source will reach its lowest luminosity at a level
set by the temperature of the neutron star's core, which 
depends on the mass accretion rate over a 10,000 year time 
scale \citep{colpi01}.

To some extent, the X-ray coverage in the 1980s limits our
knowledge of the EXO~1747--214 outburst history.  Historically,
the source has only been detected by {\em EXOSAT}, and it was
detected in 1984 June and 1985 April.  In both cases, the 
persistent X-ray flux was near $10^{-9}$ 
erg~cm$^{-2}$~s$^{-1}$.  The missions with all-sky coverage in 
that era were {\em Ariel V}, which lasted until 1980 and 
{\em Ginga} which began in 1987 \citep{csl97}.  The 90\% sky 
coverage from {\em Tenma} during 1983--1984 make it unlikely 
that the EXO~1747--214 outburst started much before 1984 June.  
Thus, assuming that EXO~1747--214 had a single outburst, its 
duration was very likely between 10 months and $\sim$3 years.
For the 13 year outburst from KS~1731--260, \cite{rutledge02}
find that the neutron star cooling is dominated by the core
rather than the crust 1 year after the outburst for a
high-conductivity crust and 30 years after the outburst for
a low-conductivity crust.  Given the shorter duration of
the EXO~1747--214 outburst, it is likely that EXO~1747--214
was in the core-dominated phase when the {\em Chandra}
observation occurred.

We can calculate the outburst recurrence time predicted by
the \cite{bbr98} theory for EXO~1747--214.  The recurrence
time is given by 
$t_{\rm rec} = (t_{\rm otb}/130)(F_{\rm otb}/F_{\rm q})$
\citep{wijnands01,tomsick04_2123}, where $t_{\rm otb}$ and
$F_{\rm otb}$ are the outburst duration and average flux
level during the outburst, respectively, and $F_{\rm q}$ 
is the quiescent flux level.  Assuming an outburst 
duration of 10 months (at the lower end of the range 
derived above) and an average outburst flux of $10^{-9}$ 
erg~cm$^{-2}$~s$^{-1}$, the upper limit on the quiescent
flux of $5\times 10^{-15}$ erg~cm$^{-2}$~s$^{-1}$
implies that $t_{\rm rec} > 1300$ years.  If, in the
future, it is found that the recurrence time is in fact
shorter than this value, then a mechanism for enhanced
cooling of the core would be necessary.  

\section{Conclusions and Future Work}

EXO~1747--214 is one of several neutron star systems with
a quiescent luminosity that is lower than the typical
$10^{32-34}$ erg~s$^{-1}$ range.  Our {\em Chandra} 
observation indicates a 0.3--8 keV luminosity upper limit 
of $L < 7\times 10^{31}$ erg~s$^{-1}$.  In the {\em Chandra}
image, there is only one source that is a possible 
EXO~1747--214 quiescent counterpart.  The source is detected 
at a significance level of 4.1-$\sigma$, leaving a small
chance that it is spurious, and it is just outside the 90\% 
confidence {\em EXOSAT} error circle.  Overall, we cannot 
rule out the possibility that the source is unrelated to 
EXO~1747--214.  Our optical and IR images show that the 
position of this possible counterpart is consistent with 
three blended optical/IR sources with $R$-band magnitudes 
between 19.4 and 19.8 and $J$-band magnitudes between 17.2 
and 17.6.  If one of these is the EXO~1747--214 counterpart, 
it would imply a distance of a few kpc, significantly lower 
than the 11 kpc upper limit derived from the peak X-ray 
burst flux.  This, in turn, would imply a significantly
lower quiescent X-ray luminosity.

As EXO~1747--214 is one of the faintest (if not the faintest 
of the) neutron star transients in quiescence, it is 
important to obtain deeper X-ray observations to determine 
the actual quiescent flux.  This is important for studies of 
neutron star cooling as well as the comparison between the 
quiescent radiative efficiency of black hole vs.~neutron star 
sources.  Another avenue for further study is to obtain 
optical spectra of the candidate counterparts.  For example,
detecting an H$\alpha$ emission line would be a strong X-ray 
binary indicator.  Finally, it is worth pointing out that 
finding the optical counterpart is likely the only way that 
we will be able to improve the determination of the source 
distance.

\acknowledgments

JAT would especially like to thank L.~Angelini for assistance 
with using the data in the {\em EXOSAT} archive and G.~Tovmassian 
for useful discussions.  JAT acknowledges partial support from 
{\em Chandra} award number GO3-4041X issued by the {\em Chandra} 
X-ray Observatory Center, which is operated by the Smithsonian 
Astrophysical Observatory for and on behalf of NASA under contract 
NAS8-03060.  Data for this work were obtained at the W.M. Keck 
Observatory, which is operated as a scientific partnership among 
the University of California, Caltech and NASA.  The SIMBAD 
database and the HEASARC Data Archive were used in preparing this 
paper.



\clearpage

\begin{table}
\caption{EXO 1747--214 X-Ray and Optical Observations\label{tab:obs}}
\begin{minipage}{\linewidth}
\footnotesize
\begin{tabular}{ccccc} \hline \hline
Start               & Observatory & Instrument & Energy & Exposure\\
Time (UT)           &  &  & Band/Filter & Time (s)\\ \hline \hline
1985 April 8, 2.3 h & {\em EXOSAT} & CMA & Thin Lexan\footnote{The
three filters cover the 0.05--2.5 keV energy range with the different
effective area curves described in the text.} & 869\\
1985 April 11, 10.0 h & {\em EXOSAT} & ME & 1--20 keV & 22,450\\
1985 April 11, 10.4 h & {\em EXOSAT} & CMA & Thin Lexan$^{a}$ & 9,171\\
1985 April 11, 11.2 h & {\em EXOSAT} & CMA & Al/Parylene$^{a}$ & 2,266\\
1985 April 11, 11.9 h & {\em EXOSAT} & CMA & Boron$^{a}$ & 2,493\\
2002 February 24, 9.5 h & CTIO\footnote{We used the 4 meter telescope
at Cerro Tololo Inter-American Observatory with the OSIRIS (Optical 
System for Imaging and Low-Resolution Integrated Spectroscopy) 
instrument.} & OSIRIS$^{b}$ & $J$ & 120\\
2003 July 31, 18.2 h & {\em Chandra} & ACIS-S & 0.3--8 keV & 24,479\\
2003 September 21, 5.9 h & Keck & ESI & $R$ & 600\\ \hline
\end{tabular}
\end{minipage}
\end{table}

\end{document}